\documentclass[conference]{IEEEtran}
\usepackage[left=0.653in,right=0.640in,top=0.75in,bottom=1.067in]{geometry}
\usepackage{cite}
\usepackage{algorithm,algpseudocode}
\usepackage{enumitem}
\usepackage{amsmath}
\usepackage{amsmath,amssymb,amsfonts}
\usepackage{graphicx}
\usepackage{textcomp}
\usepackage{xcolor}
\usepackage{mathtools}
\usepackage{multirow}
\usepackage{booktabs}
\usepackage{xcolor} %
\newcommand*\subtxt[1]{_{\textnormal{#1}}}

\DeclareMathOperator*{\argmin}{\arg\min}   

\def\BibTeX{{\rm B\kern-.05em{\sc i\kern-.025em b}\kern-.08em
		T\kern-.1667em\lower.7ex\hbox{E}\kern-.125emX}}
\begin{document}
	
	\title{	SwinLSTM Autoencoder for Temporal-Spatial-Frequency Domain CSI Compression in Massive MIMO Systems\\
	}
	
	\author{\IEEEauthorblockN{Aakash Saini$^{\ast\ddagger}$, Yunchou Xing$^\dagger$, Jee Hyun Kim$^\ast$, Amir Ahmadian Tehrani$^\ast$, Wolfgang Gerstacker$^\ddagger$}
		\IEEEauthorblockA{
			$^\ast$\textit{Nokia Standards, Munich, Germany}\\
				$^\dagger$\textit{Nokia Standards, Naperville, IL, USA}\\
		$^\ddagger$\textit{Institute for Digital Communications, Friedrich-Alexander-Universität
			Erlangen-Nürnberg, Erlangen, Germany}}
	}

	\maketitle
	
	\begin{abstract}
		This study presents a parameter-light, low-complexity artificial intelligence/machine learning (AI/ML) model that enhances channel state information (CSI) feedback in wireless systems by jointly exploiting temporal, spatial, and frequency (TSF) domain correlations. While traditional frameworks use autoencoder for CSI compression at the user equipment (UE) and reconstruction at the network (NW) side in spatial-frequency (SF), massive multiple-input multiple-output (mMIMO) systems in low mobility scenarios exhibit strong temporal correlation alongside frequency and spatial correlations. An autoencoder architecture alone is insufficient to exploit the TSF domain correlation in CSI; a recurrent element is also required. To address the vanishing gradients problem, researchers in recent works propose state-of-the-art TSF domain CSI compression architectures that combine recurrent networks for temporal correlation exploitation with deep pre-trained autoencoders that handle SF domain CSI compression. However, this approach increases the number of parameters and computational complexity. To jointly utilize correlations across the TSF domain, we propose a novel, parameter-light, low-complexity AI/ML-based recurrent autoencoder architecture to compress CSI at the UE side and reconstruct it on the NW side while minimizing CSI feedback overhead.		
	\end{abstract}
	
	
	\section{Introduction}
	The mMIMO systems are key components in 5G and beyond\cite{b1}. Base stations (BSs) with numerous antennas are able to provide high-speed connectivity without interference in principle \cite{b3}. Time division duplex (TDD) and frequency division duplex (FDD) transmission modes of mMIMO systems require precise, real-time CSI for optimal performance. While TDD mode systems can leverage channel reciprocity to simplify CSI acquisition, FDD mode systems demand accurate and instantaneous CSI through an uplink CSI feedback mechanism \cite{b4}. As the dimensions of antenna arrays grow, so does the overhead for transmitting CSI, consuming valuable uplink resources. This necessitates efficient compression and sensing of time-critical CSI.
	
	CSI feedback approaches can be categorized as explicit, or implicit CSI. \cite{bXplicitImplicit}. Explicit CSI refers to raw channel information without assumptions about transmission schemes, while implicit CSI incorporates specific assumptions about transmission and reception. Explicit CSI provides comprehensive channel information but requires higher uplink bandwidth, whereas implicit CSI offers more accurate link adaptation with reduced uplink resource consumption. This distinction significantly impacts system performance in practical deployments.
	
	Early deep learning approaches like CsiNet \cite{b5} focused primarily on compressing explicit CSI by deploying a deep learning-based encoder at the UE to compress the CSI and a corresponding decoder at the NW to reconstruct the CSI. This autoencoder-based CSI feedback enhancement is referred to as a two-sided model approach, as the components of the autoencoder are present on both the UE side and the NW side\cite{twoSided}. Authors in \cite{b2} offer a thorough review of these methods, examining architectures, challenges, and future research prospects. For SF domain CSI compression, transformer-based models have demonstrated superior performance, though their complexity prompted the development of localized self-attention-based architectures \cite{9926175} and shifted window multi-head self-attention designs \cite{10419637} to address computational demands.
	
	Research has subsequently evolved toward exploiting the SF domain correlation of implicit CSI \cite{biImCsiNET}, which provides a more efficient representation. Building upon this foundation, researchers in \cite{b6,b7} extended SF domain compression to incorporate temporal correlations, enabling TSF domain CSI compression and prediction. These state-of-the-art approaches employ deep pretrained modules, to address the vanishing gradient problem caused by long computational paths from input to output. However, this modular approach introduces significant computational overhead. For instance, using the same CSI dimensions as in this work, the pretrained module Bidirectional LSTM-based Encoder Network (BEN) in \cite{b6} alone contains approximately 734,016 parameters, contributing to a total architecture size of up to approximately 57.8 million parameters.
	 An alternative approach in \cite{jintao} proposes joint CSI compression and prediction utilizing temporal correlation via ConvLSTM. Research in \cite{CATT_Tdoc} demonstrates that integrating ConvLSTM layers with transformer-based encoder-decoder pairs (ConvLSTM-TF) achieves significant performance improvements. Although ConvLSTM-TF reduces dependence on pretrained autoencoder weights, its architecture remains challenging for UE implementation.
 	
	Recently, SwinLSTM has been introduced as a novel recurrent architecture designed for efficient spatiotemporal representation extraction, primarily targeted to computer vision tasks \cite{swinLSTM}. The main contributions of this paper are:
	\begin{itemize}
		\item We propose a parameter-light, low-complexity deep learning model architecture referred to as \textit{SLATE} (SwinLSTM AutoEncoder) that applies the SwinLSTM cell \cite{swinLSTM} to efficiently compress and reconstruct CSI matrices by jointly exploiting TSF domain correlations.
		
		\item The results of system-level simulations confirm that our proposed TSF compression architecture, \textit{SLATE}, outperforms both the Release 16 enhanced TypeII codebook and transformer-based SF domain while achieving comparable performance to ConvLSTM-TF-based TSF domain CSI compression methods with approximately 76\% fewer parameters (700K vs 3.0M) and 86\% lower complexity.
	\end{itemize}
	 \begin{figure*}[t!]
	 	\centerline{\includegraphics[width=0.98\textwidth]{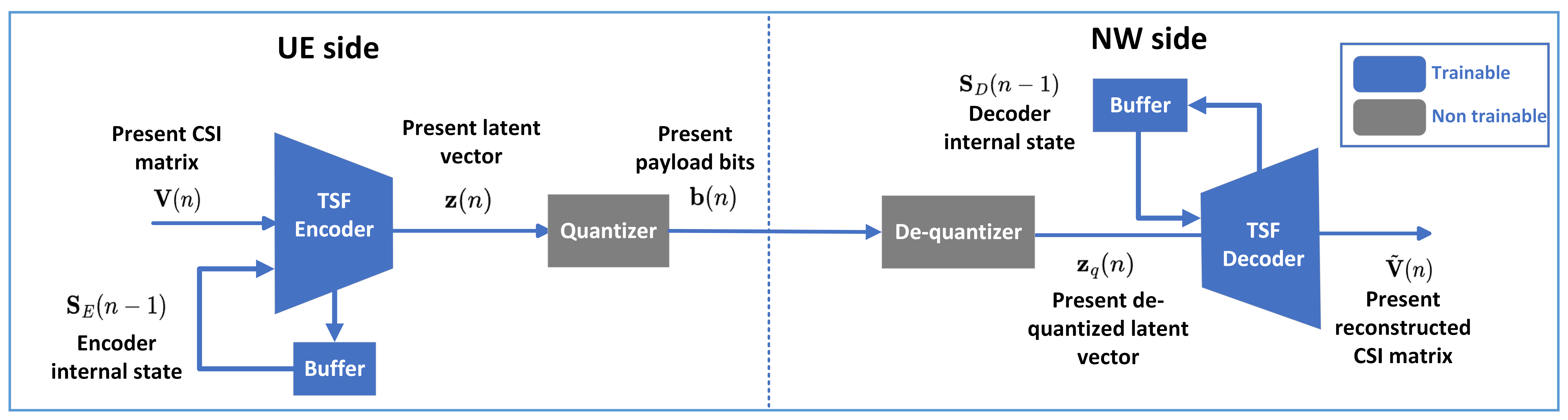}}
	 	\caption{AI/ML-driven temporal-spatial-frequency (TSF) domain CSI compression framework.}
	 	\label{fig1}
	 \end{figure*} 
      The paper is organized as follows. Section \ref{section2} presents the system model. Section \ref{section4} elucidates our proposed SLATE architecture. Section \ref{section5} presents and discusses simulation results. Finally, Section \ref{section6} concludes the paper and outlines future research directions.
 	 
	\section{System model} \label{section2}
	We extend our previous system model, which studied training strategies for two-sided model approach for CSI compression \cite{10615736}, to study TSF domain CSI compression while retaining key notations. The framework considers an FDD-based mMIMO-OFDM system serving users from the set $\mathcal{K} = \left\{1,\cdots, K\right\}$. Each BS employs $N\subtxt{Tx}$ dual-polarized antennas, while the $k$-th UE utilizes $N_{\text{Rx}, k}$ dual-polarized antennas such that $N\subtxt{Tx} \gg N_{\text{Rx}, k}$. The BS transmits data streams with indices $\mathcal{S}=\left\{1,2,\cdots,\Lambda\right\}$ across allocated frequency resources $\mathcal{F}$, where $|\mathcal{S}|=\Lambda\leq N\subtxt{Tx}$. Each UE receives a subset of streams $\mathcal{S}_{k} \subseteq \mathcal{S}$ on its assigned frequency resources $\mathcal{F}_k \subseteq \mathcal{F}$, with $|\mathcal{S}_{k}|=\Lambda_k \leq N_{\text{Rx}, k}$. The complex data vector for the $f$-th resource block (RB) across streams is denoted as $\mathbf{x}_{f}(n) = \left[x_{1,f}(n), x_{2,f}(n),...,x_{\Lambda,f}(n)\right]^{T} \in \mathbb{C}^{\Lambda}$. For the $k$-th UE, stream $s \in \mathcal{S}_k$, and RB $f\in \mathcal{F}_k$, the received symbol $\tilde{y}_{k,s,f}(n) \in \mathbb{C}$ is expressed as
	\begin{equation}\label{eq:rxSym}
		\begin{split}
			\tilde{y}_{k,s,f} (n)= \mathbf{f}^{H}_{k,s,f}(n)\mathbf{H}_{k,f}(n)\mathbf{v}_{s,f}(n)x_{s,f}(n) + \\
			\sum_{ \substack{j \in \mathcal{S}\\ j \neq s }}\mathbf{f}^{H}_{k,s,f}(n)\mathbf{H}_{k,f}(n)\mathbf{v}_{j,f}(n)x_{j,f}(n) + \tilde{n}_{k,f}(n),
		\end{split}
	\end{equation}
	where $\mathbf{f}_{k, s, f}(n) \in \mathbb{C}^{N\subtxt{Rx,k}}$ and $\mathbf{v}_{s,f}(n) \in \mathbb{C}^{N\subtxt{Tx}}$ denote the combining and precoding vectors, respectively, for the $f$-th RB. Here, the MIMO channel matrix between BS and $k$-th user for the $f$-th RB is represented by $\mathbf{H}_{k,f}(n) \in \mathbb{C}^{N\subtxt{Rx,k} \times N\subtxt{Tx}}$. After combining, the received signal is subject to a noise term $\tilde{n}_{k,f}(n):= \mathbf{f}^{H}_{k,s,f}(n)\mathbf{n}_{k,f}(n)$, where $\mathbf{n}_{k,f}(n) \in \mathbb{C}^{N\subtxt{Rx}}$ follows a circularly symmetric complex Gaussian distribution $\mathcal{CN}(0,\sigma_{k,f}^2 \mathbf{I}_{{N\subtxt{Rx,k}}})$.
	
	Following the 3rd Generation Partnership Project (3GPP) specifications, we combine every four RBs from a total of 56 RBs into one subband, indexed by $f_{\text{SB}}$ ranging from $1$ to $N_{\text{SB}}=14$, to reduce the uplink feedback overhead. For the $k$-th UE on the $f_{\text{SB}}$-th subband, assuming a linear solution, the optimal precoders correspond to the dominant channel eigenvectors of $\mathbf{H}^H_{k,f_{\text{SB}}}(n)\mathbf{H}_{k,f_\text{SB}}(n)$ \cite{1468321}. The implicit CSI matrix feedback from the $k$-th UE to the BS comprises these dominant channel eigenvectors as $\mathbf{V}_{k}(n) = \left[\mathbf v_1^k(n), \mathbf v_2^k(n), \ldots, \mathbf v^k_{N_{\text{SB}}}(n)\right] \in \mathbb{C}^{N_{\text{Tx}} \times N_{\text{SB}}}$. 
	
	For temporal analysis, $N$ successive precoder matrices at a given time sample $n$ are concatenated as $\hat{\mathbf{V}}_{k} = \left[\mathbf{V}_{k}(N), \cdots, \mathbf{V}_{k}(n), \mathbf{V}_{k}(n-1), \cdots, \mathbf{V}_{k}(1)\right] \in \mathbb{R}^{N\times 2 \times N_{\subtxt{Tx}} \times N_{\subtxt{SB}}}$, with real and imaginary components represented by the factor of 2. Limited-rate uplink channels restrict full complex floating-point implicit CSI feedback. Therefore, the system employs $K$ AI/ML-based TSF encoder-decoder pairs, one per user, trained with UE-specific training data $\mathcal{V} = \left\{\hat{\mathbf{V}}_{1}, \ldots \hat{\mathbf{V}}_{K}\right\}$. Here, each UE's training data set $\hat{\mathbf{V}}_{k} \in \mathbb{R}^{M\times N\times 2\times  N_{\text{Tx}}\times N_{\text{SB}}}$ contains $M$ training samples across $N$ time instances.
	
	Figure \ref{fig1} depicts the CSI feedback framework tailored for TSF compression. The TSF encoder processes current CSI matrix $\mathbf{V}(n)$, producing latent vector $\mathbf{z}(n)$ while updating its previous internal state $\mathbf{S}_E(n-1)$. The quantized latent vector is presented by payload bits $\mathbf{b}(n)$. On the NW side, de-quantized latent vector $\mathbf{z}_q(n)$ enables CSI matrix reconstruction $\mathbf{\tilde{V}}(n)$ through the TSF decoder while storing the decoder's previous internal state $\mathbf{S}_D(n-1)$ in the buffer. Both encoder and decoder employ buffers to utilize past internal states, enabling the exploitation of temporal correlation. The set $\mathcal{Z} = \left\{\mathbf{Z}^1, \cdots, \mathbf{Z}^K\right\}$ contains time-correlated floating-point latent vectors, where $\mathbf{Z}^k = \left[\mathbf{z}^k(N),\cdots,\mathbf{z}^k(n),\mathbf{z}^k(n-1),\cdots, \mathbf{z}^k(1)\right] \in \mathbb{R}^{L_{dim}\times N}$ represents $N$ SF domain CSI samples for the $k$-th UE. Functions $f_{\mathcal{Q}}(\cdot): \mathcal{Z} \to \mathbb{Z}^{B}_{2}$ and $f_{\mathcal{Q'}}(\cdot): \mathbb{Z}^{B}_{2} \to \mathcal{Z}_q$ perform quantization and de-quantization of $\mathbf{z}(n)$ to $L_{dim}\times B$ bits and vice versa, with $B$ bits resolution. The de-quantized latent vector for the $k$-th UE, at $n$-th time sample, is given as
	\begin{equation}\label{eq:Q-dQ}
		\mathbf{z}^k_q(n) = f_{\mathcal{Q'}}\left(f_{\mathcal{Q}}\left(\mathbf{z}^k(n)\right)\right..
	\end{equation}
	
	The collection of time dependent de-quantized latent vectors across $K$ users is represented as $\mathcal{Z}_q = \left\{\mathbf{Z}^1_q,\cdots, \mathbf{Z}^K_q\right\}$, where $\mathbf{Z}^k_q = \left[\mathbf{z}^k_q(N),\cdots,\mathbf{z}^k_q(n),\mathbf{z}^k_q(n-1),\cdots, \mathbf{z}^k_q(1)\right] \in \mathbb{R}^{L_{dim}\times N}$ denotes $N$ time dependent de-quantized latent vectors for the $k$-th UE. The hypothesis space $\mathcal{H}$ contains all potential functions that can be represented by the chosen model architecture and parameters, which may include but are not limited to correlation relationships in the data. For the $k$-th UE, the encoding function $f_{\mathcal{E}_k}\left(\mathbf{V}_{k}(n), \mathbf{S}_{E}(n-1); \Theta_{\mathrm{E}_k}\right)\in \mathcal{H}$ maps $\mathcal{V} \to \mathcal{Z}$, where $\Theta_{\mathrm{E}_k}$ represents the TSF encoder's trainable parameters. We assume perfect CSI acquisition at the BS, neglecting uplink transmission errors. The corresponding network-side decoding function $f_{\mathcal{D}_k}\left(\mathbf{z}^{k}_{q}(n), \mathbf{S}_D(n-1); \Theta_{\mathrm{D}_k}\right)\in \mathcal{H}$ maps $\mathcal{Z}_q \to \mathcal{V}$, with $\Theta_{\mathrm{D}_k}$ denoting the TSF decoder's trainable parameters. The current reconstructed CSI at the BS side is given as
	\begin{equation}\label{reconstructedCSI}
	\begin{split}
		\tilde{\mathbf{V}}_k(n) = f_{\mathcal{D}_k}\left(f_{\mathcal{Q'}}\left(f_{\mathcal{Q}}\left(f_{\mathcal{E}_k}\left(\mathbf{V}_{k}(n),\mathbf{S}_{E}(n-1);\right. \right. \right. \right.\\ \left. \left. \left. \left. \Theta_{\mathrm{E}_k}\right)\right)\right), \mathbf{S}_{D}(n-1); \Theta_{\mathrm{D}_{k}}\right).
	\end{split}
	\end{equation}
	
	In order to exploit the temporal correlation, $N$ successive reconstructed CSI matrices are stacked as $$\mathbf{V}^{\prime}_{k} = \left[\tilde{\mathbf{V}}_{k}(N), \tilde{\mathbf{V}}_{k}(N-1), \cdots, \tilde{\mathbf{V}}_{k}(1)\right] \in \mathbb{C}^{N\times N_{\subtxt{Tx}} \times N_{\subtxt{SB}}}.$$ In contrast to the separate training approaches for the UE-side encoder and NW-side decoder discussed in \cite{10615736}, we implement joint, offline training of the encoder-decoder pair. The optimization problem for training the $k$-th TSF encoder-decoder pair is accordingly formulated as
	\begin{equation}\label{jointTraining}
		\left\{\hat{\Theta}_{\mathrm{E}_{k}}, \hat{\Theta}_{\mathrm{D}_{k}}\right\} = \argmin_{\Theta_{\mathrm{E}_{k}}\Theta_{\mathrm{D}_{k}}} \mathbb{E}_{\hat{\mathbf{V}}_k} \left[\ell\left(\hat{\mathbf{V}}_k, \mathbf{V}^{\prime}_k\right)\right],
	\end{equation}
	where $\ell(\cdot)$ is the selected loss function and $\mathbb{E}_{\hat{\mathbf{V}}_{k}}$ denotes expectation over training labels $\hat{\mathbf{V}}_{k}$. The negative SGCS, averaged across time samples and subbands, defines the loss as
	\begin{equation}\label{eq:SGCS}
		\ell\left(\hat{\mathbf{V}}_{k}, \mathbf{V}^{\prime}_k\right) =- \frac{1}{N}\frac{1}{N_{\text{SB}}}\sum_{n=1}^{N}\sum_{\mathnormal{f}_{\text{SB}}=1}^{N_{\text{SB}}}\frac{\left\|\mathbf{v}_{\mathnormal{f}_{\text{SB}}}^{k}(n)^H  \tilde{\mathbf{v}}_{\mathnormal{f}_{\text{SB}}}^{k}(n)\right\|^2}{\left\| \mathbf{v}_{\mathnormal{f}_{\text{SB}}}^{k}(n)\right\|^{2} \left\| \tilde{\mathbf{v}}_{\mathnormal{f}_{\text{SB}}}^{k}(n)\right\|^{2}},
	\end{equation}
	where the $\mathnormal{f}_{\text{SB}}$-th column of $\mathbf{V}_{k}(n)$, $\mathbf{v}_{\mathnormal{f}_{\text{SB}}}^{k}(n)$, contains the time-varying dominant channel eigenvectors for subband $\mathnormal{f}_{\text{SB}}$ of UE $k$, $\tilde{\mathbf{v}}_{\mathnormal{f}_{\text{SB}}}^{k}(n)$ is its reconstruction, and $\|\cdot\|$ denotes the Euclidean norm.
     \begin{table*}[t!]
    	\centering
     	\caption{Details of the proposed SLATE architecture}
     	\begin{tabular}{c|l|c|c|c}
     		\hline\hline
     		& Layer name & Output shape & Activation & Number of trainable parameters  \\
     		&  &  & Function &   \\
     		\hline
     		\multirow{4}{*}{UE} & Input & $2 \times N_{\text{Tx}} \times N_\text{SB}$ &  & 0  \\
     		& Patch Embedding & $ \frac{N_\text{SB}N_\text{Tx}}{P_hP_w}\times E_{dim}$ & LReLU & $\left(2P_hP_w+3\right)E_{dim}$  \\
     		& $r$-th Patch Merge & $\frac{N_{\text{SB}}N_{\text{Tx}}}{P_hP_w \left(\prod_{i=r}^1 d_i\right)^2} \times \left(\prod_{i=r}^1 d_i\right)E_{dim}$ &  & $4d_rd_{r-1}^2E_{dim}^2 + 8E_{dim}: d_0 = 1$  \\
     		& $r$-th SwinLSTM Cell & $\frac{N_{\text{SB}}N_{\text{Tx}}}{P_hP_w \left(\prod_{i=r}^1 d_i\right)^2} \times \left(\prod_{i=r}^1 d_i\right)E_{dim}$ &  & $14\alpha_r\left( \prod_{i=r}^1 d_i\right)^2E_{dim}^2 + 14\alpha_r\left(\prod_{i=r}^1 d_i\right)E_{dim}$  \\
     		& MLP & $L_{dim}$ & Tanh & $\left(2+L_{dim}\right)\frac{N_\text{Tx}N_\text{SB}E_{dim}}{P_hP_w\left(\prod_{i=r}^1 d_i\right)}+L_{dim}$  \\
     		\hline
     		\multirow{4}{*}{NW} & MLP & $\frac{N_{\text{SB}}N_{\text{Tx}}}{P_hP_w \left(\prod_{i=r^\prime}^1 u_i\right)^2} \times \left(\prod_{i=r^\prime}^1 u_i\right)E_{dim}:$ &  & $\left(1+L_{dim}\right)\frac{N_\text{Tx}N_\text{SB}E_{dim}}{P_hP_w\left(\prod_{i=r^\prime}^1 u_i\right)} +2L_{dim} $  \\
     			& &$ r^\prime=R-r+1$ & & \\
     		& $r$-th SwinLSTM Cell & $\frac{N_{\text{SB}}N_{\text{Tx}}}{P_hP_w \left(\prod_{i=r^\prime}^1 u_i\right)^2} \times \left(\prod_{i=r^\prime}^1 u_i\right)E_{dim}$ &  & $14\alpha_r\left(\prod_{i=r^\prime}^1 u_i\right)^2E_{dim}^2 +14\alpha_r\left(\prod_{i=r^\prime}^1 u_i\right)E_{dim}$\\
     		& $r$-th Patch Expand & $\frac{N_{\text{SB}}N_{\text{Tx}}}{P_hP_w \left(\prod_{i=r^\prime-1}^1 u_i\right)^2} \times \left(\prod_{i=r^\prime-1}^1 u_i\right)E_{dim}$ & LReLU & $2\left(\prod_{i=r^\prime}^1 u_i\right)^2E_{dim}^2+\left(\prod_{i=r^\prime}^1 u_i\right)E_{dim}$ \\
     		& Patch Extract & $2 \times N_{\text{Tx}} \times N_\text{SB}$ &  & $2\left(P_hP_wE_{dim}+1\right)$\\
     		& Output & $2 \times N_{\text{Tx}} \times N_\text{SB}$ &  & 0  \\
     		\hline\hline
     	\end{tabular}
     	\label{table:3}
     \end{table*}

    \section{Proposed SLATE (SwinLSTM AutoEncoder) Architecture for TSF domain CSI compression} \label{section4}
\begin{figure*}[t!]
    	\centerline{\includegraphics[width=0.98\textwidth]{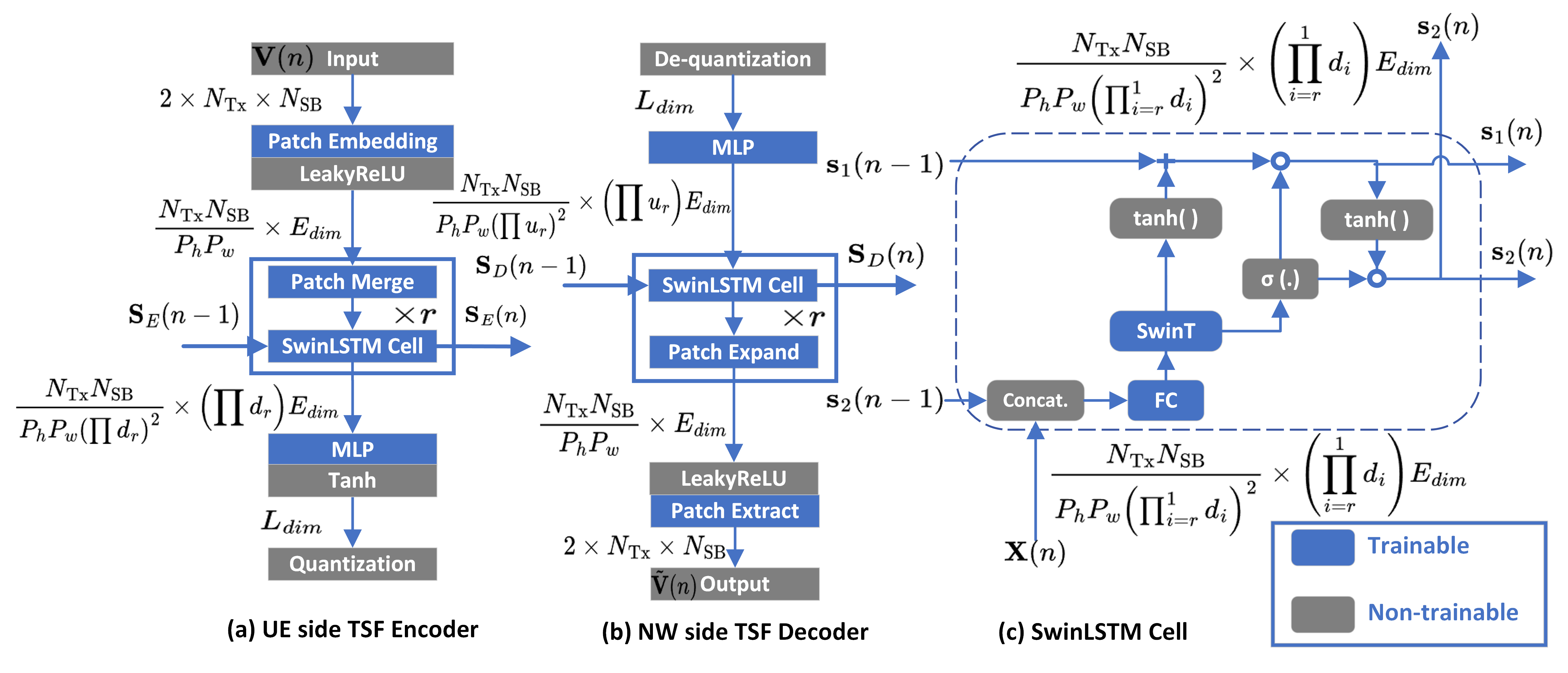}}
    	\caption{Architecture of the proposed SLATE-based Temporal-Spatial-Frequency (TSF) domain CSI compression framework consisting of (a) UE-side TSF encoder, (b) NW-side TSF decoder, and (c) SwinLSTM cell in TSF encoder integrating shifted window transformer blocks for joint spatial-frequency correlation learning and LSTM components for temporal dependency modeling of CSI.}
    	\label{fig3}
    \end{figure*}

Fig. \ref{fig3} illustrates the architecture of the proposed SLATE-based TSF domain CSI compression. The real and imaginary parts of $\mathbf{V}(n) \in \mathbb{R}^{2\times N_{\text{Tx}} \times N_{\text{SB}}}$ are stacked to form the spatial-frequency (SF) domain CSI for the $n$-th time sample. As shown in Fig. \ref{fig3} (a), this SF domain CSI matrix is input to the UE-side TSF encoder. A patch embedding layer divides the CSI into patches based on the hyperparameters patch height ($P_h$) and patch width ($P_w$). It assigns a vector of length $E_{dim}$ to each patch, resulting in dimensions $\frac{N_{\text{Tx}} N_{\text{SB}}}{P_hP_w}\times E_{dim}$. This is followed by a leaky rectified linear unit (LReLU) activation function. The architecture then comprises repetitions of patch merge and SwinLSTM cell layers, denoted by the index $r\in \left\{1,2,\cdots, R\right\}$. The patch merge layers reduce spatial and frequency resolution by a downscaling factor $d_r$ while increasing the embedding dimension by the same factor $d_r$. The SwinLSTM cell for the $r$-th layer, combining shifted window (Swin) transformers of depth $\alpha_r$ and LSTM gate, extracts spatial-frequency features and captures temporal correlations using previous state $\mathbf{S}_{E}\left(n-1\right)$. After number of $r$ repetitions, output dimensions are $\frac{N_{\text{Tx}} N_{\text{SB}}}{P_hP_w\left(\prod_{i=r}^1 d_i\right)^2}\times E_{dim}\left(\prod_{i=r}^1 d_i\right)$. A multi-layer perceptron (MLP) head with Tanh non-linearity creates a bottleneck, compressing the CSI attention map to a latent vector of dimension $L_{dim}$. Finally, 2-bit uniform scalar quantization is applied to each dimension of the latent vectors. The CSI feedback report consists of a binary mapping of the quantized latent vector. For a rank $\Lambda_k$ transmission in the proposed scheme, the payload overhead in terms of the number of bits in each CSI report can be determined as \begin{equation}\label{eq:Overhead} N_{\text{O}} = 2 \cdot L_{dim} \cdot \Lambda_k. \end{equation}
   	\begin{figure}
    	\centerline{\includegraphics[width=0.40\textwidth]{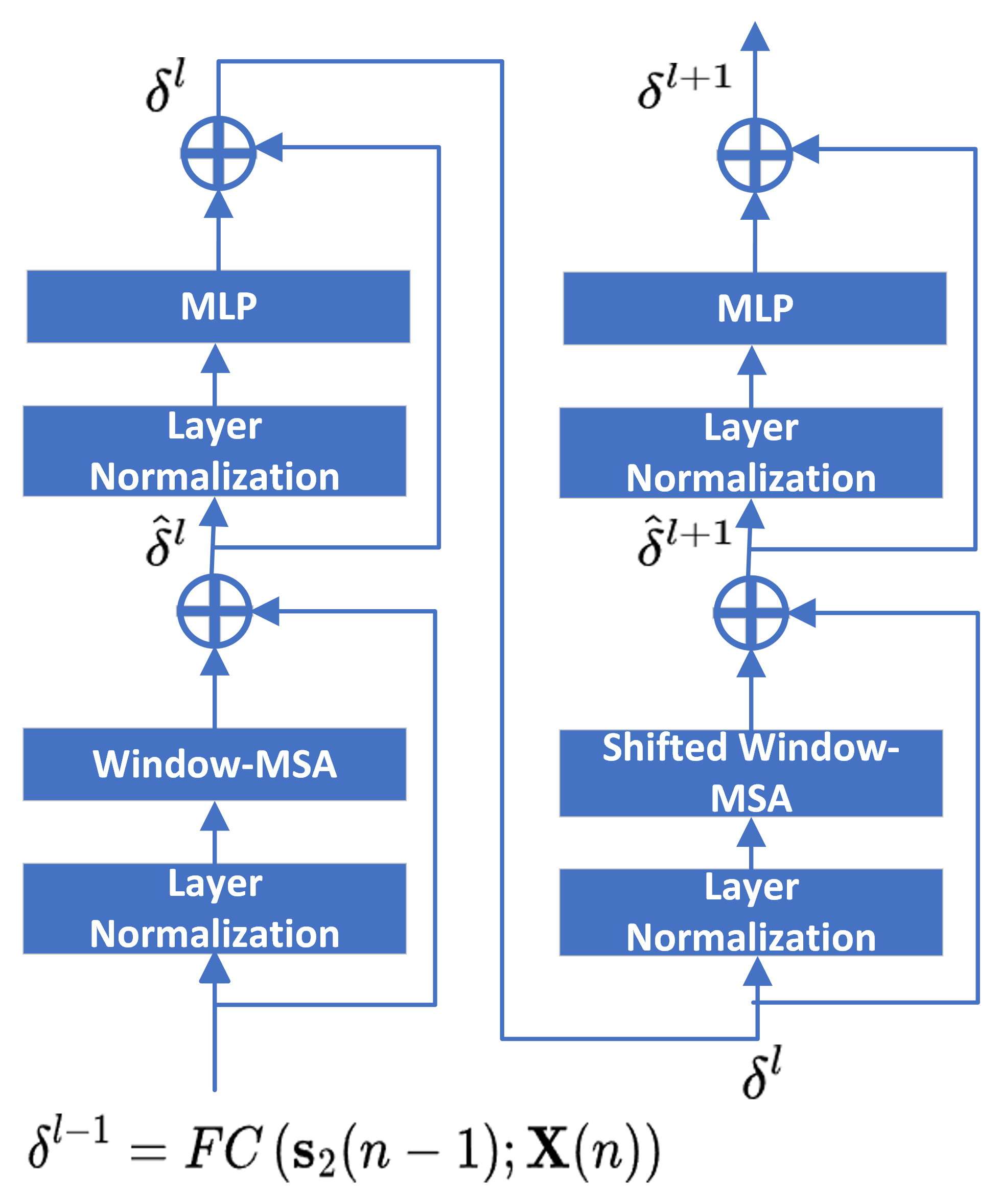}}
    	\caption{Shifted window (Swin) transformer for the $l$-th layer.}
    	\label{fig4}
    \end{figure}
    \begin{figure}
    	\centerline{\includegraphics[width=0.45\textwidth]{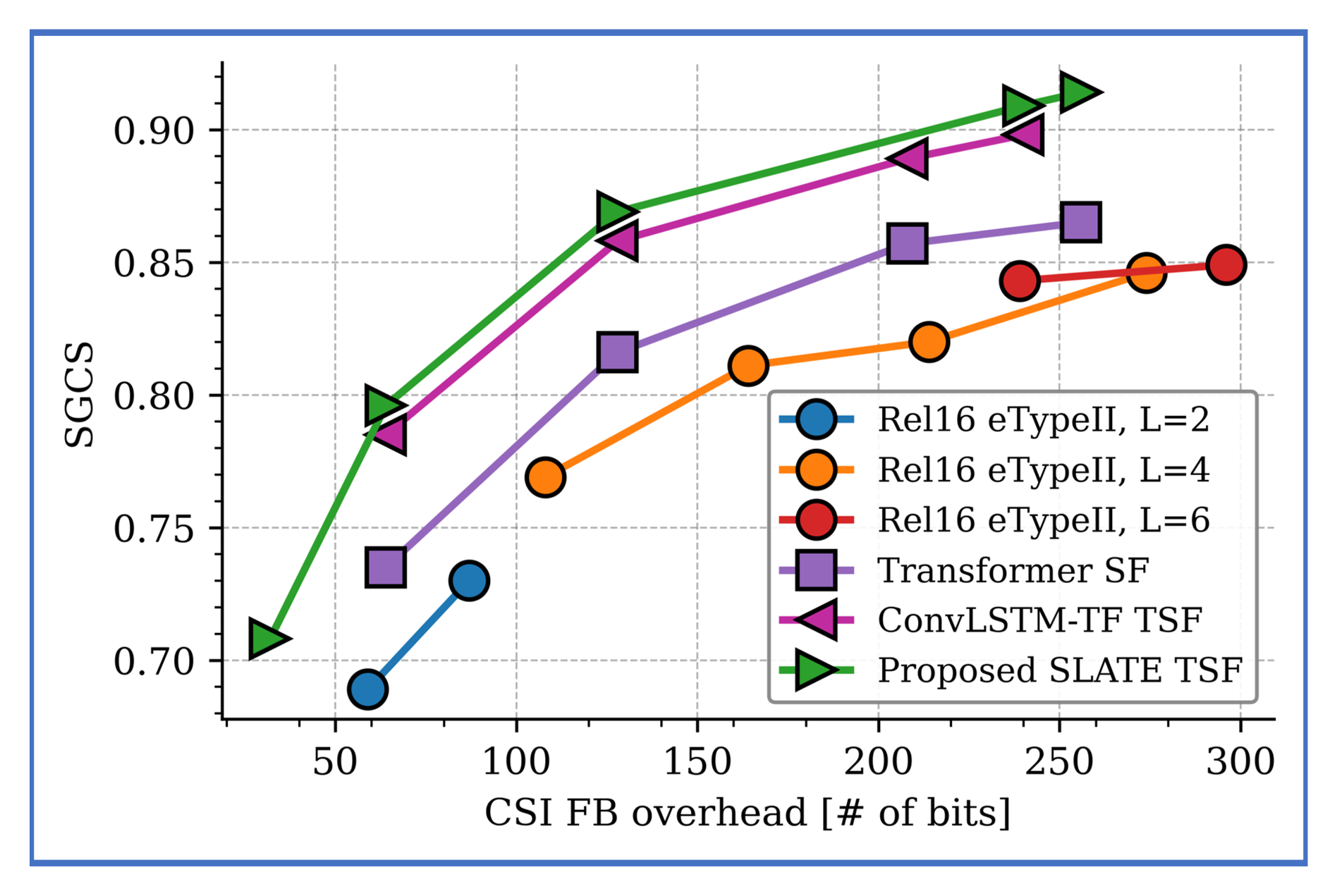}}
    	\caption{SGCS as a function of maximum number of overhead bits, for the baseline scheme (``Rel16 eTypeII''), the transformer-based SF domain (``Transformer SF''), the convolutional LSTM and transformer-based TSF domain (``ConvLSTM-TF TSF''), and the proposed SLATE-based TSF domain (``SLATE TSF'') CSI feedback compression scheme.}
    	\label{fig5}
    \end{figure}
     On the NW-side (Fig. \ref{fig3} (b)), a de-quantization block maps the received CSI payload bits to a de-quantized latent vector of dimension $L_{dim}$. An MLP head up-projects this vector to dimension $\frac{N_{\text{Tx}} N_{\text{SB}}}{P_hP_w\left(\prod_{i=r^\prime}^1 u_i\right)^2}\times E_{dim}\left(\prod_{i=r^\prime}^1 u_i\right): r^\prime = R-r+1$, where $u_{r^\prime}$ is the required up-scaling factor corresponding to the $r$-th layer down-scaling factor $d_r$. The decoder then applies $r$ iterations of SwinLSTM cells and patch-expand layers, mirroring the UE-side encoder structure. The SwinLSTM cell reconstructs temporal, spatial, and frequency domain attention maps using the previous decoder state $\mathbf{S}_{D}(n-1)$ and the current up-projected latent vector. Patch expanding layers increase the spatial and frequency resolution according to $u_r$ while decreasing the embedding dimensions by the same factor $u_r$. After $r$ iterations corresponding to $r$ layers, the output dimensions become $\frac{N_{\text{Tx}} N_{\text{SB}}}{P_hP_w}\times E_{dim}$. Finally, the expanded attention maps undergo LReLU non-linearity and patch extraction (2D transposed convolution), resulting in an output dimension of  $2\times N_{\text{Tx}}\times N_{\text{SB}}$. As compression and reconstruction occur independently per transmission layer, the reconstructed channel eigenvectors may lose their orthogonality, which is originally maintained in the UE-side encoder. Re-orthogonalization could, therefore, be implemented to preserve eigenvector orthogonality across the transmission layers.
     \begin{figure*}[t!]
     	\centerline{\includegraphics[width=0.95\textwidth]{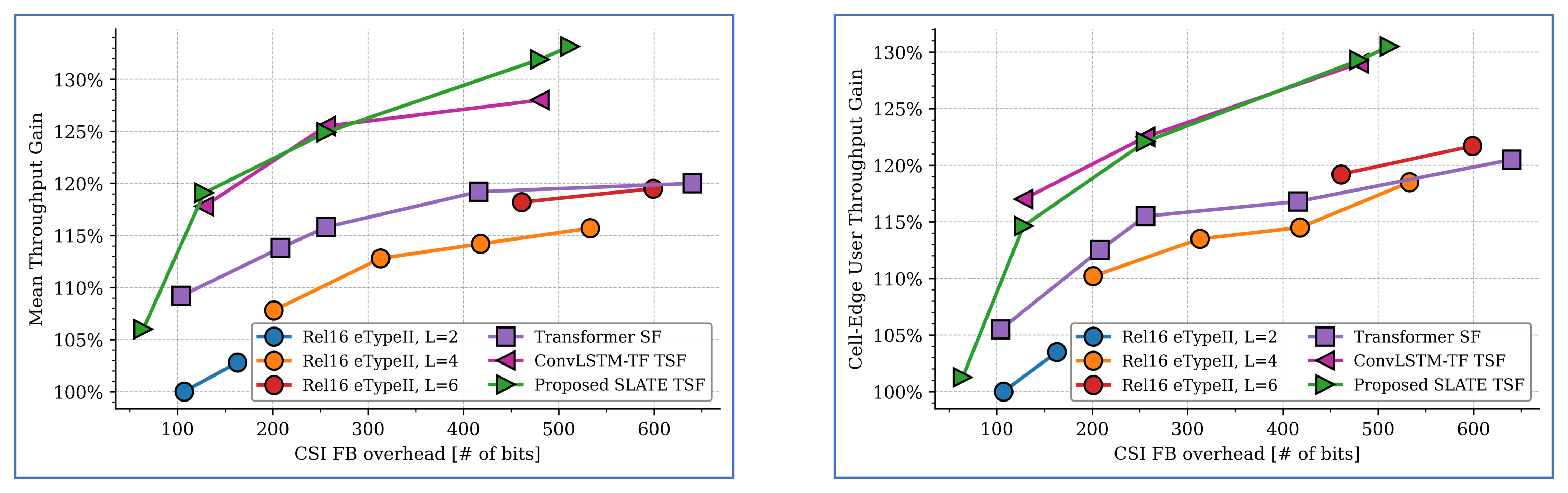}}
     	\caption{User Throughput Gain (left) and Cell-Edge User Throughput Gain (right) for rank-2 transmission, shown as a function of maximum number of overhead bits, for the baseline scheme (``Rel16 eTypeII''), the Transformer-based SF domain (``Transformer SF''), the convolutional LSTM and transformer-based TSF domain (``ConvLSTM-TF TSF''), and the proposed SLATE-based TSF domain (``SLATE TSF'') CSI feedback scheme.}
     	\label{fig6}
     \end{figure*}

Fig. \ref{fig3} (c) illustrates the SwinLSTM cell \cite{swinLSTM} structure employed for the UE side TSF encoder. Internal state components $\mathbf{s}_1(n)$ and $\mathbf{s}_2(n)$ capture short and long-term temporal correlations, which are updated in each time step $n$. The Swin transformer block \cite{SwinTrans} of depth $\alpha =2$, shown in Fig. \ref{fig4}, learns spatial and frequency domain correlations. This approach reduces the quadratic computational complexity of global multi-head self-attention (MSA) in vision transformers by using window-based and shifted window-based MSA to nearly linear, depending upon the window size. The shifted window transformer comprises two consecutive layers that are repeated:
\begin{enumerate}
	\item The $l$-th layer processes input $\delta^{l-1}$ through layer normalization, window-MSA, and an MLP.
	\item The second layer applies layer normalization, shifted window-MSA, and another MLP.
\end{enumerate}
The shifted window mechanism in the second layer interconnects different windows, creating a global attention map while maintaining computational efficiency. Table~\ref{table:3} presents the architectural specification of the proposed SLATE-based TSF compression model divided into UE side TSF encoder and NW side TSF decoder components, detailing each layer's name, output shape, activation function, and the number of trainable parameters. Moreover, the architecture follows a symmetric pattern with patch-based processing and SwinLSTM cells at both ends.

    \section{Performance Evaluation} \label{section5}
    \begin{algorithm}[t!]
    	\caption{Offline joint training of the proposed SLATE-based TSF encoder and decoder}
    	\label{alg:training}
    	\begin{algorithmic}[1]
    		
    		\Require \begin{itemize}[noitemsep,topsep=0pt]
    			\item[$\alpha$:] Learning rate
    			\item[$N_B$:] Batch size
    			\item[$E$:] Number of training epochs
    			\item[$N$:] Number of temporal samples in input CSI sequence
    		\end{itemize}
    		
    		
    		\State \textbf{Initialize:} Parameters of $k^{\text{th}}$ encoder-decoder $\left\{\Theta_{\mathrm{E}_k},\Theta_{\mathrm{D}_k}\right\}$
    		
    		\For{epoch $i = 1$ to $E$}
    		\For{each batch}
    		\State \textbf{Sample batch:} 
    		$\displaystyle \Big\{[\mathbf{V}_k(1), \cdots, \mathbf{V}_k(N)]^{(b)}\Big\}_{b=1}^{N_B}$ 
    		
    		\State \textbf{Initialize internal states:}
    		\Statex \hspace{2.75em} $\mathbf{S}(0) = [\mathbf{s}_1(0), \mathbf{s}_2(0)] \hspace{0.25em} : \hspace{0.25em} \mathbf{S}_E(n) = \mathbf{S}_D(n) = \mathbf{S}(n)$
    		
    		\For{$n = 1$ to $N$}
    		\State \textbf{Update:} Get the present reconstructed CSI and 
    		\Statex \hspace{4.2em} internal states:
    		\Statex \hspace{4.2em} $\tilde{\mathbf{V}}_k(n), \mathbf{S}(n)$ using Eq.~\eqref{reconstructedCSI}
    		\EndFor
    		\State  $ \ell \leftarrow \ell\left(\left[\hat{\mathbf{V}}_k(1): \hat{\mathbf{V}}_k(N)\right],\left[\tilde{\mathbf{V}}_k(1): \tilde{\mathbf{V}}_k(N)\right]\right)$
    		\Statex \hspace{2.75em} using Eq.~\eqref{eq:SGCS} 
    		\State Update $ \Theta_k \leftarrow \Theta_k - \alpha\nabla_{\Theta_k}\ell \hspace{0.25em} : \hspace{0.25em}\Theta_k=\left\{\Theta_{\mathrm{E}_k},\Theta_{\mathrm{D}_k}\right\}$
    		\EndFor
    		\EndFor		
    	\end{algorithmic}
    \end{algorithm}
    We evaluate the performance of the proposed \textit{SLATE} based TSF domain CSI compression scheme using the training dataset described in Section~\ref{section2} and compare its performance to that of transformer-based SF and convLSTM-TF-based TSF, respectively. Matrix $\hat{\mathbf{V}}$ is constructed by stacking $N=10$ SF domain CSI samples using a 3GPP fifth generation (5G) new radio (NR) system-level simulator, which generates both training and test datasets and performs system-level simulations using the parameters outlined in TABLE \ref{Tb:simulationConfig}. The adopted dataset comprises 168000 training and 42000 test samples. The hyperparameters are set as follows: $P_h=1$, $P_w=4$, $E_{dim}=32$, with patch merge and SwinLSTM cell repetition factor $R=3$. The downscaling factors ($d_r$) are chosen to 1, 1, and 2, while Swin transformer depths ($\alpha_r$) in SwinLSTM cells are selected to 2, 4, and 2 for the first, second, and third layers, respectively, on both encoder and decoder sides. We vary $L_{dim}$ (16, 32, 64, 120, 128) to achieve a different number of payload overhead bits as per Eq.~\eqref{eq:Overhead}. The NW-side decoder up-scaling factors ($u_r$) are set to 2, 1, and 1 for the three patch extract layers. The model is implemented and trained with PyTorch. The model training uses Adam optimizer with learning rate $\alpha=0.0001$, batch size $N_B=128$, number of epochs $E=650$, and number of temporal samples $N=10$. The UE-side TSF encoder and NW-side TSF decoder are trained jointly offline as detailed in Algorithm~\ref{alg:training}. Training is performed in a quantization-aware manner with non-trainable quantization and de-quantization blocks positioned between the encoder and decoder. Due to the non-differentiable nature of these blocks, their gradients are set to 1 during backpropagation.
    
    In Fig.~\ref{fig5}, the SGCS is shown as a function of CSI feedback overhead (in number of bits) for different methods. Here, Release-16 enhanced TypeII codebook (non-AI/ML) \cite{eTypeIIRel16}, transformer-based SF domain, convLSTM-TF-based TSF domain, and our proposed SLATE-based TSF domain CSI compression approach are considered. The results indicate that SLATE-based TSF domain CSI compression maintains the same performance level as the 3GPP Release 16 eTypeII \cite{eTypeIIRel16} (with spatial domain beams L=2,4,6) while reducing the CSI feedback overhead by 50-175 bits per CSI report. The proposed approach achieves similar performance to the transformer-based SF domain compression model while reducing overhead by 25-120 bits per CSI report. Compared to the convLSTM-TF-based TSF domain compression model, it maintains comparable performance at a similar number of overhead bits while requiring 76\% fewer parameters and reducing computational complexity by 86\% in terms of floating point operations, as shown in Table~\ref{table:4}. It can also be noticed that the proposed architecture for TSF domain CSI compression outperforms the transformer-based SF domain compression while exhibiting a significantly reduced number of parameters. 
        \begin{table} [t!]
    	\renewcommand{\arraystretch}{1.31125}
    	\caption{System-level simulation parameters}
    	\label{Tb:simulationConfig}
    	\centering
    	\setlength\tabcolsep{4pt} 
    	\begin{tabular}{r l}
    		\toprule
    		Scenario       & 3GPP TR 38.901 UMa\\
    		Channel Drops & 10  \\
    		Numerology &$f$\textsubscript{DL}= 4.0 GHz,\\
    		& $\Delta f$ = 30 KHz\\ 
    		Bandwidth & 20 MHz \\
    		Num. cell sites & 7 sites\\
    		Sectors per site & 3 sectors per site\\
    		\midrule
    		Number of UEs & 10 per sector\\
    		BS antenna height & 25 m\\
    		Distributions of UE (Indoor\%, Outdoor\%) &(80, 20)\\
    		BS antenna    &($M^\prime = 2, N^\prime = 8, P = 2$),\\
    		& $N\subtxt{Tx} = 32$,\\
    		& $d$\textsubscript{H}= $0.5\lambda$, $d$\textsubscript{V}= $0.8\lambda$\\
    		UE antenna & ($M^\prime = 1, N^\prime = 2, P = 2)$,\\
    		& $N\subtxt{Rx} = 4$, $d$\textsubscript{H}= $0.5\lambda$  \\
    		Number of subbands  & $N_{\text{SB}} = 14$ \\
    		Number of time samples  & $N = 10$ \\
    		\midrule
    		UE velocity (Indoor, Outdoor)  & (3, 30) km/h \\
    		Feedback schemes & Rel-16 eTypeII\\
    		& AI/ML-based (autoencoder)\\
    		Feedback interval & $T$\textsubscript{CSI} = 5 ms \\
    		Quantization  & Uniform fixed scalar\\
    		Quantization bits  & 2 bits per latent\\
    		&vector dimension\\
    		\bottomrule
    	\end{tabular}
    \end{table}
   
    After offline model training, we integrated the trained weights into our system-level simulator. We evaluated the proposed CSI compression scheme's performance through MU-MIMO 3GPP NR (5G) system-level simulations. The simulations measured mean user throughput and mean cell-edge user throughput across 210 UEs, comprising 80\% indoor users moving at 3 km/h and 20\% outdoor users moving at 30 km/h. We used ten channel realizations, each running for the duration of at least 2000 ms radio transmission. Fig. \ref{fig6} (left) shows the average user throughput gain versus maximum CSI feedback overhead, with 95\% of UEs maintaining a rank indicator of 2. The proposed \textit{SLATE} based TSF domain compression architecture demonstrates superior performance compared to the existing methods: a throughput gain of 7.5-15\% over the 3GPP Release-16 eTypeII codebook-based method \cite{eTypeIIRel16}, 7.5-10\% over transformer-based SF domain compression, and comparable performance to ConvLSTM-TF-based TSF domain compression at significantly reduced number of parameters and FLOPs (Table~\ref{table:4}). Fig. \ref{fig6} (right) validates these findings by showing cell-edge performance, indicating a mean cell-edge user throughput gain of 10-15\% over 3GPP Release-16 eTypeII \cite{eTypeIIRel16}, 7-12\% gain over transformer-based SF domain compression, and equivalent performance to ConvLSTM-TF-based at improved computational efficiency as detailed in Table~\ref{table:4}.
        \begin{table}[t!]
    	\centering
    	\caption{Approximate number of parameters and FLOPs for CSI compression schemes (${L_{dim}=64}$, ${N_{\text{Tx}}=32}$, ${N_{\text{SB}}=14}$) }
    	\begin{tabular}{c|l|c|c}
    		\hline\hline
    		& Architecture & Parameter (\textbf{M}) & FLOPs/$T_{\text{CSI}}$ (\textbf{M})\\
    		\hline
    	    & Transformer SF & 2.8 & 30.7 \\
    		& ConvLSTM-TF TSF & 3.0 & 358.0 \\
    		& Proposed SLATE TSF &  \textbf{0.7} & \textbf{48.5}  \\
    		\hline\hline
    	\end{tabular}
    	\label{table:4}
    \end{table}
    \section{Conclusion} \label{section6}
    In this paper, a novel, parameter-light, and computationally efficient recurrent autoencoder framework that incorporates SwinLSTM cells to compress CSI in the TSF domain has been introduced. Our architecture effectively leverages the correlations across temporal, spatial, and frequency dimensions to achieve CSI compression and reconstruction while minimizing the number of overhead bits. Through comprehensive MU-MIMO system-level simulations, we have demonstrated the significant performance advantages of our SLATE-based TSF compression approach.
    Future research directions include evaluating the robustness of the architecture across varying mobility conditions, analyzing the impact of CSI estimation errors, and optimizing the model architecture to achieve a further optimized balance between computational efficiency and performance.
	
    \section*{Acknowledgment}
    Part of this work has been performed within the Horizon Europe project CENTRIC (101096379), funded by the European Union. The part of the research utilizes and adapts source code from S. Tang \textit{et al.} \cite{swinLSTM} under the MIT License. The authors acknowledge their colleagues' contributions to developing Nokia's internal system-level simulator and Nokia Machine Learning Platform (NMLP).
     
	\bibliographystyle{ieeetr}
	\bibliography{MyBib} 	

\begin{thebibliography}{10}

\bibitem{b1}
O.~Elijah {\em et~al.}, ``Intelligent massive {MIMO} systems for beyond {5G}
  networks: An overview and future trends,'' {\em IEEE Access}, vol.~10,
  pp.~102532--102563, 2022.

\bibitem{b3}
A.~Ashikhmin, L.~Li, and T.~L. Marzetta, ``Interference reduction in multi-cell
  massive {MIMO} systems with large-scale fading precoding,'' {\em IEEE
  Transactions on Information Theory}, vol.~64, no.~9, pp.~6340--6361, 2018.

\bibitem{b4}
E.~Nayebi and B.~D. Rao, ``Semi-blind channel estimation for multiuser massive
  {MIMO} systems,'' {\em IEEE Transactions on Signal Processing}, vol.~66,
  no.~2, pp.~540--553, 2018.

\bibitem{bXplicitImplicit}
B.~Clerckx, G.~Kim, J.~Choi, and Y.-J. Hong, ``Explicit vs. implicit feedback
  for su and mu-mimo,'' in {\em 2010 IEEE Global Telecommunications Conference
  GLOBECOM 2010}, pp.~1--5, 2010.

\bibitem{b5}
C.-K. Wen, W.-T. Shih, and S.~Jin, ``Deep learning for massive {MIMO} {CSI}
  feedback,'' {\em IEEE Wireless Communications Letters}, vol.~7, no.~5,
  pp.~748--751, 2018.

\bibitem{twoSided}
J.~Guo, C.-K. Wen, S.~Jin, and X.~Li, ``{AI} for {CSI} feedback enhancement in
  {5G}-advanced,'' {\em IEEE Wireless Communications}, vol.~31, no.~3,
  pp.~169--176, 2024.

\bibitem{b2}
J.~Guo {\em et~al.}, ``Overview of deep learning-based {CSI} feedback in
  massive {MIMO} systems,'' {\em IEEE Transactions on Communications}, vol.~70,
  no.~12, pp.~8017--8045, 2022.

\bibitem{9926175}
S.~Mourya, S.~Amuru, and K.~K. Kuchi, ``A spatially separable attention
  mechanism for massive {MIMO} {CSI} feedback,'' {\em IEEE Wireless
  Communications Letters}, vol.~12, no.~1, pp.~40--44, 2023.

\bibitem{10419637}
J.~Cheng {\em et~al.}, ``Swin transformer-based {CSI} feedback for massive
  {MIMO},'' in {\em 2023 IEEE 23rd International Conference on Communication
  Technology (ICCT)}, pp.~809--814, 2023.

\bibitem{biImCsiNET}
M.~Chen {\em et~al.}, ``Deep learning-based implicit {CSI} feedback in massive
  {MIMO},'' {\em IEEE Transactions on Communications}, vol.~70, no.~2,
  pp.~935--950, 2022.

\bibitem{b6}
C.~Jiang {\em et~al.}, ``Deep learning-based implicit {CSI} feedback for
  time-varying massive {MIMO} channels,'' in {\em ICC 2023 - IEEE International
  Conference on Communications}, pp.~4955--4960, 2023.

\bibitem{b7}
S.~Kadambar {\em et~al.}, ``Deep learning based joint {CSI} compression and
  prediction for beyond-{5G} systems,'' in {\em GLOBECOM 2023 - 2023 IEEE
  Global Communications Conference}, pp.~4792--4797, 2023.

\bibitem{jintao}
Z.~Ren, X.~Zhang, and J.~Wang, ``Joint {CSI} feedback and prediction with deep
  learning in high-speed scenarios,'' in {\em 2024 IEEE/CIC International
  Conference on Communications in China (ICCC)}, pp.~1910--1915, 2024.

\bibitem{CATT_Tdoc}
CATT, ``{R1-2406356; Further study on AI/ML for {CSI} compression},'' Tech.
  Rep., 3GPP TSG RAN WG1 \#118, August 2024.

\bibitem{swinLSTM}
S.~Tang, C.~Li, P.~Zhang, and R.~Tang, ``{SwinLSTM}: Improving spatiotemporal
  prediction accuracy using swin transformer and {LSTM},'' in {\em 2023
  IEEE/CVF International Conference on Computer Vision (ICCV)},
  pp.~13424--13433, 2023.

\bibitem{10615736}
A.~Saini {\em et~al.}, ``Network-first separate training with raw dataset
  sharing: A training approach for {AI/ML}-driven {CSI} feedback,'' in {\em
  2024 IEEE International Conference on Communications Workshops (ICC
  Workshops)}, pp.~1950--1955, 2024.

\bibitem{1468321}
D.~Love and R.~Heath, ``Limited feedback unitary precoding for spatial
  multiplexing systems,'' {\em IEEE Transactions on Information Theory},
  vol.~51, no.~8, pp.~2967--2976, 2005.

\bibitem{SwinTrans}
Z.~Liu {\em et~al.}, ``Swin transformer: Hierarchical vision transformer using
  shifted windows,'' {\em CoRR}, vol.~abs/2103.14030, 2021.

\bibitem{eTypeIIRel16}
3GPP, ``{TS 138 214 - V16.2.0 - {5G}; NR; Physical layer procedures for data
  (3GPP TS 38.214 version 16.2.0 Release 16)},'' Tech. Rep., 2020.

\end{thebibliography}
\end{document}